# The neural and cognitive architecture for learning from a small sample

**Aurelio Cortese[1], Benedetto De Martino[2,3], Mitsuo Kawato[1,4]**

1. Computational Neuroscience Laboratories, ATR Institute International, Kyoto, Japan

2. Institute of Cognitive Neuroscience, University College of London, Alexandra House, 17-19 Queen Square, London WC1N 3AR, UK

3. Wellcome Centre for Human Neuroimaging, University College London, WC1N 3BG London, United Kingdom

4. RIKEN Center for Advanced Intelligence Project, ATR Institute International, Kyoto, Japan

**Abstract**

Artificial intelligence algorithms are capable of fantastic exploits, yet they are still grossly inefficient compared with the brain's ability to learn from few exemplars or solve problems that have not been explicitly defined. What is the secret that the evolution of human intelligence has unlocked? Generalization is one answer, but there is more to it. The brain does not directly solve difficult problems, it is able to recast them into new and more tractable problems. Here we propose a model whereby higher cognitive functions profoundly interact with reinforcement learning to drastically reduce the degrees of freedom of the search space, simplifying complex problems and fostering more efficient learning.

**Introduction**

Artificial Intelligence (AI) has come a long way since the summer of 1956, when it was first envisaged at the Dartmouth Summer Research Project on Artificial Intelligence. In the last ten years we have witnessed how the principles of supervised and reinforcement learning, when embedded in neural networks composed of many hidden layers ('deep neural networks', or 'DNN'), can reach - and often surpass - human-level performances in visual object recognition, and in playing video-games and GO [1–3]. Despite DNN's massive computational capabilities, there are two aspects that temper these accomplishments: first, the number of training samples required to reach acceptable performances is huge - tens or hundreds of millions; second, these architectures show a limited ability to generalize to new tasks/settings that were not encountered during training. These limitations become largely evident in motor control as shown by the clumsy behaviour of humanoid robots in [the DARPA robotic challenge](#) [4].

A second avenue of exciting progress in AI has come from probabilistic machine learning (a.k.a. Bayesian machine learning) [5], where agents can achieve impressive performance on one-shot learning or using a limited amount of examples [6,7]. This probabilistic approach resonates well with intuitive theories of human cognitive development and inductive reasoning [8]. The learning algorithm tries to find among all possible models the one that best explains the data (or, by extension, infer what causes the reality perceived through the sensorium). This approach, while conceptually appealing, is unlikely to provide a realistic model of how the brain operates. The main issue is that fully probabilistic inference might work well in simple and well constrained conditions, but becomes quickly computationally intractable for more complex and unconstrained scenarios. To exacerbate the problem, in order to function efficiently, probabilistic programs have to be endowed with ad-hoc definitions of the necessary representations [5]. Strictly speaking, in Bayesian inference we usually do not have a principled way to select initial priors. Generalization is thus limited to the class of problems for which the program was designed for [6,7].

A considerable hurdle for artificial agents concerns generalization; how can machine learning algorithms deal with new and never-experienced scenarios? Humans and animals can easily and appropriately respond to new scenarios, mostly transferring knowledge acquired in loosely related contexts. What are the brain mechanisms that enable the human brain with its remarkable generalization capacity? Plain reinforcement learning is too slow, and hierarchical architectures [9], albeit ameliorating the algorithm by subdividing learning among multiple systems and meta-variables [10,11], remain dependent on the need for ad-hoc definitions. Here we suggest that brains do not simply solve supervised classification problems but transform them into different - and more tractable - problems. We propose that an adaptive role of higher cognition is to allow precisely this transformation to take place.

More specifically, we propose a model of how higher cognition is able to simultaneously operate the dimensionality reduction and feature selection processes necessary for simplifying complex problems. Adjusting the degree of synchronization between neurons has been suggested as one possible way to control the degrees-of freedom of a neural system [12]. We draw on similarities with simulated annealing, exploring how different frequency modes of brain dynamics serve as inherent implementation channels to reduce degrees-of-freedom and reach optimal solutions.

**Computational advantages of higher cognitive functions in learning**

Statistical learning theory of singular problems demonstrates that the generalization error is given by dividing the degrees-of-freedom ($d$) of the search space by the number of training sample ($n$): $e \propto d/(2n)$ [13,14]. If brains ($d \sim 10^{11}$ neurons) need to solve arbitrary classification problems utilizing only a few hundred learning samples ($n \sim 10^2$), the generalization error would become huge, at least $10^{11}/10^2 = 10^9$. We postulate that brains transform these intractable learning problems into more feasible reinforcement learning problems with small degrees of freedom while being guided by reward and penalty. Higher cognitive functions such as attention, memory, concept formation, and metacognition might find low-dimensional manifolds of meta-representations that are essential for learning from a small sample (fig. 1). Here we will briefly review findings from attention, memory, concept formation, and metacognition, focusing on their role in facilitating learning. We are aware that these are vast and active areas of investigation and that it would be hard to do justice to all the relevant work that has been done. We have therefore decided to provide a snapshot of properties, modules and architectures that we believe are particularly relevant to inspire the development of new AI architectures. It is important to recognize that recent work in the field of machine learning has also started to incorporate some of the intuitions discussed here.

Attention is the ability to direct computing resources toward relevant dimensions (stimulus attributes, spatial location, etc.) for focal processing, acting as a filter to amplify relevant information while dampening background clutter [15]. But how does an agent learn what to attend? Rewards and punishments serve to constrain attentional focus [16], and attending to specific features rather than to the whole improves versatility [17]. Essentially, we learn *what* to attend to at the same time as we are paying attention to what we are learning [18,19]. In machine learning, a useful and efficient way to use attention mechanisms is to decompose tasks or questions into a series of simpler operations [20], or target specific parts of a query (e.g. particular words in a sentence) [21].

An intelligent agent must also be able to remember or even sometime forget past events. Accordingly, episodic memory plays a special role in goal-directed behavior and learning [22]. Reality is statistically structured however, and forms of gist-like memory can enhance reinforcement learning [23]. Schematic memory still depends on episodes; it is by virtue of statistics over individual traces that summaries can be created. Not surprisingly, both persistence (remembering) and transience (forgetting) are essential ingredients to optimize decision-making [24]. Human-like memory processes are very different from what is usually considered in AI agents, where memory is often deterministic and non-sparse as well as storing all information. Linking neural networks to external buffer memory resources already produces impressive learning capabilities, unattainable by classic neural networks architectures [25,26]. The development of predictive memory architectures, where memory formation itself is guided by a process of predictive inference [27], is one step towards systems storing only relevant information.

Concepts *are* abstractions, closely intertwined with schematic memories. Concepts can be created almost at will, and a key aspect is that they can be connected, creating conceptual maps [28]. Being highly hierarchical and compositional, more abstract concepts can be formed from existing ones. New concepts or conceptual maps can emerge from learning, but can also direct subsequent learning [29]. Concepts share obvious links with memory in their ability to represent schematized information, but work in AI has yet to follow this line of thought. Conceptual representations in AI are currently restricted to simple visual domain examples that make use of the principles of hierarchy and compositionality [30].

Self-monitoring processes, a more abstract class of cognitive functions, can encompass much richer representations. The ability to monitor one's thoughts is referred to as metacognition, and is linked to the psychological construct of confidence, i.e. how good an agent is at keeping track of the probability of a choice being correct [31,32]. This aspect is very important for AI since it dovetails with a broad range of phenomena such as error monitoring and reality checking [33]. Of particular relevance to AI systems is the ability to explicitly track the evolution of the level of self-knowledge, which might provide biological agents with significant advantages when interacting with their environment [34–38]. Although metacognition and consciousness are intimately related, the question of what is the computational advantage of consciousness itself remains currently unanswered. Consciousness could represent the selection of information for global broadcasting within the system, making it flexibly available for local (and distant) computational units [33]. In machine learning consciousness could also be interpreted as a powerful constraint on low-dimensional representations [39]. Earlier efforts suggest that some forms of self-monitoring are computationally simple and can directly arise even in two-layer attractor networks [40]. Generative adversarial networks (GANs) are an exciting development in this direction: a generative model captures the data distribution, and a discriminative model, akin to metacognition, operates a reality check on new samples [41].

**Neural implementation of high level cognitive architecture**

To generate solutions leading to efficient learning and flexible behaviors, nature had to solve physical constraints. The brain cannot be equipped with ad hoc representations for every possible event in the world, since the horizon of possible states is practically infinite; moreover, it does not have unlimited computing resources. These constraints may be inspirational for developing new AI, yet it is important to keep in mind that some biological constraints (e.g. positive neuronal firing rates) may be bypassed by in-silico intelligent systems.

In its most basic interpretation, solving complex problems for the brain accounts to finding the relevant (hidden) states for RL. One solution to accelerate the search for hidden states is to capitalize on the brain's massively parallelized neural circuit architecture. Parallel searches are instantiated in multiple recurrent circuits linking basal ganglia with the cortex (Fig. 2). These recurrent circuits effectively are information-transmitting loops: they can carry task-dependent explicit representations (stimuli, goals, etc.), abstract summaries, reward prediction errors (RPEs), predicted states. Although they carry heterogeneous information, parallel loops do not function independently from each other. Rather, loops formed by sparse neural populations continuously interact at the synaptic level through cooperation and competition. Excitatory interactions (cooperation) appear between loops with similar, inclusive and related representations. In contrast, inhibitory interactions (competition) develop between loops with exclusive, different or unrelated representations. Due to the dynamical nature of the neural network comprising these excitatory and inhibitory interactions, a winner-take-all scenario emerges [42,43]. That is, only the loop with the "best" representation survives while other loops are suppressed. Here "best" means the loop associated with the representation that minimizes RPE. Therefore, selection of the best loop essentially corresponds to the automatic selection of relevant states for RL.

Excitatory and inhibitory interactions can occur virtually anywhere in the brain. However, the basal ganglia should play the most important role in these synaptic interactions for the following reasons. Because of multiple inhibitions and direct, indirect, hyperdirect pathways linking basal ganglia to cortex, winner-take-all computations can best be implemented in basal ganglia [43,44]. Furthermore, RPEs are largely computed in basal ganglia [45], making these nuclei the ideal focal point for RPE-based loop comparison and selection.

So far, we have discussed a relatively simplified model that is amenable to clearly delineate the theory. The reality of the brain is nevertheless more intricate. Several brain areas are likely interacting to orchestrate an efficient search and ensure convergence to task-relevant low-dimensional manifolds. Prefrontal, sensorimotor, hippocampal cortices, as well as cerebellum, thalamus, and basal ganglia all make loops. Above this automatic machinery, what is the role of higher cognitive functions? How can they further accelerate learning computations? Metacognition, attention and memory synchronize abstract representations in prefrontal cortex (PFC) or hippocampal formation (HPC) with concrete representations in sensorimotor areas. Recurrent connections between these regions connect reinforcement learning mechanisms with representational and abstraction engines that makes for an ideal candidate circuitry (Fig. 2).

Dopamine inputs to the HPC ensure that learned or partially learned rules are conceptualized and stored in memory [46]. But HPC function stretches far beyond memories, and one influential idea is that it plays a key role in building cognitive maps for spatial [47] and conceptual navigation [28]. HPC neurons functionality probably extends so that they take the role of predictive units extracting structure and low-dimensional bases of the world [48,49]. These discoveries are in line with a recent proposal that during decision making, inferable state-to-state transitions represented in the cortex keep track of the evolving hidden space to accelerate learning [50,51]. More specifically, the PFC is thought to hold an exclusive position along the hierarchy of representations as the substrate forging meta-representations [11,31] and abstraction processes [10,52,53]. Furthermore, the PFC oscillatory frequencies act as mediators of abstraction: the synchronization frequency helps demix the abstraction level encoded in different regions of the PFC [54]. In fact, this is not only the case for the PFC; oscillatory frequencies effectively form communication channels throughout the brain [54–56].

At the neuronal level, representing distributions of stochastic variables in population activities is a mechanism for finding serial correlations between meta-representations and RPE. The cortex could perform probabilistic inference on such distributions either by sampling over neural populations [57], or by weighting correlations between neurons [58].

Learning new problems would invariably start with a consistent scenario: ignition of myriads of parallel loops resulting in widespread neural activity over extensive cortico-striatal networks. The selection of RL states starts with broad sweeps to evolve in a fine search. Initially, broad brain regions are equally activated and participate in the search. Next, dimension reduction and feature selection begin to drop-out less activated loops, accelerated by higher cognition (Fig. 2). Finally, only a small number of loops remain and neural activity should be concentrated to basal ganglia and the few cortical locations carrying the most relevant representations. A useful analogy for this process is simulated annealing and Gibbs sampling, optimization techniques to approximate global solutions in large search spaces [59,60]. Annealing starts by first using high temperatures causing large changes in the objective function, then iteratively descending to lower temperatures causing ever smaller rearrangements - until convergence. That is, high temperatures are a form of dimension reduction, while low temperatures are akin to feature selection. We suggest that dimensionality reduction relates to abstraction, operating at low oscillatory frequency modes with low spatial resolution and using large neural populations, while feature selection relates to specific content utilizing high frequency modes and sparse neural ensembles (Fig. 3). Low frequency synchronization delineates the horizon of relevant *dimensions* so that high frequency-based *feature* selection can happen - the key element is that in the brain these take place simultaneously, accelerating interaction and winner-take-all convergence of loop drop-out. Recent work has elegantly linked the brain's structural connectivity (particularly the thalamo-cortical system) with neural activity patterns and dynamics, providing a formal basis for harmonic patterns of certain frequencies [61]. The authors of this work demonstrate that these connectome-specific harmonics patterns self-organize through the interplay of neural excitation and inhibition in coupled dynamical systems.

At last, we can now delineate how cognitive functions may affect and expedite learning processes. We have a system composed of massively parallelized modules centered around RL machinery, where communication frequency determines the abstraction level of representations, and where cognitive functions have the ability to synchronize representations at different abstraction levels. Initially, when the search is still in its infancy and characterized by activity over broad areas, the system is typified by low abstraction, and low spatiotemporal-frequency synchronization, with most processes unconscious. RPEs may be tied to any aspects of the task, with most RPEs being unspecific and irrelevant. Attention and memory play important roles at this stage for the selection of relevant features (i.e., loops) [62] through synchronization in high spatiotemporal-frequency channels [55,63]. As learning progresses, feature-specific RPEs become predominant, and the number of activated loops greatly decreases. The abstraction level increases because feature-specific RPEs can be represented summarily, hence further reducing the dimensionality and complexity of the problem. Hidden states in reinforcement learning can now be discovered more readily because the search domain has shrunk. The degree of certainty or uncertainty on neural meta-representations can provide a fast track to which states are relevant or irrelevant in reinforcement learning, by virtue of its self-monitoring property [33].

Metacognition should thus plays a central role when learning switches from initial broad search to rule acquisition in localized sparse neural ensembles. Furthermore, learning may reach the highest level of abstraction by piercing the veil of consciousness. Conscious representation of rules can be interpreted as a maximally abstract summary, a tensor with very low dimensionality that nevertheless carries fundamental information [39]. These sort of meta-representation vectors are extremely useful because they can be easily applied to new, similar but previously unexperienced, problems.

**Conclusions**

Fruitful interactions between neuroscience and AI have opened up a new exciting era beyond DNN, which require huge training samples. Brains utilize higher cognitive functions such as attention, memory, concept formation, and metacognition to transform seemingly intractable supervised learning problems with astronomical degrees-of-freedom state spaces and small samples, into reasonable reinforcement learning problems within a low-dimension meta-representation manifold.

We postulated that the neuronal mechanism implementing this transformation of computational problems is likely comprised of parallel search of low-dimensional meta-representations via synchronization of multiple loops formed by the cerebral cortex, HPC and basal ganglia. Brain connectivity and nonlinear neural dynamics provide harmonic modes spanning from low to high spatiotemporal frequencies. Interactions between different modes may provide dimension reduction and feature selection analogous to simulated annealing, albeit much faster. That is, low frequency mode could allow for dimension reduction analogous to high temperature in annealing, while high frequency modes could select a small number of features analogous to low temperature. Furthermore, real-time interactions between high and low frequency modes may enable fast parallel searches to quickly determine the reinforcement learning search domain. Attention and episodic memory are presumed mechanisms operating feature selection, while conceptualization mainly takes the form of dimension reduction. Discovery of relevant hidden states may be greatly accelerated by metacognition through synchronization of meta-representations.

Taken together these cognitive modules, acquired over millions of years by natural selection, might inspire a new generation of AI architectures that will take us one step closer to human level intelligence.

**Conflict of interest statement**

None declared.

**Acknowledgements**

A.C. and M.K. were supported by the New Energy and Industrial Technology Development Organization (NEDO) and the ImPACT Program of Council for Science, Technology and Innovation, Cabinet Office, Government of Japan; A.C. was supported by ERATO project of Japan; B.D.M. was supported by the Wellcome Trust and Royal Society (Sir Henry Dale Fellowship 102612/A/13/Z).

**FIGURES**

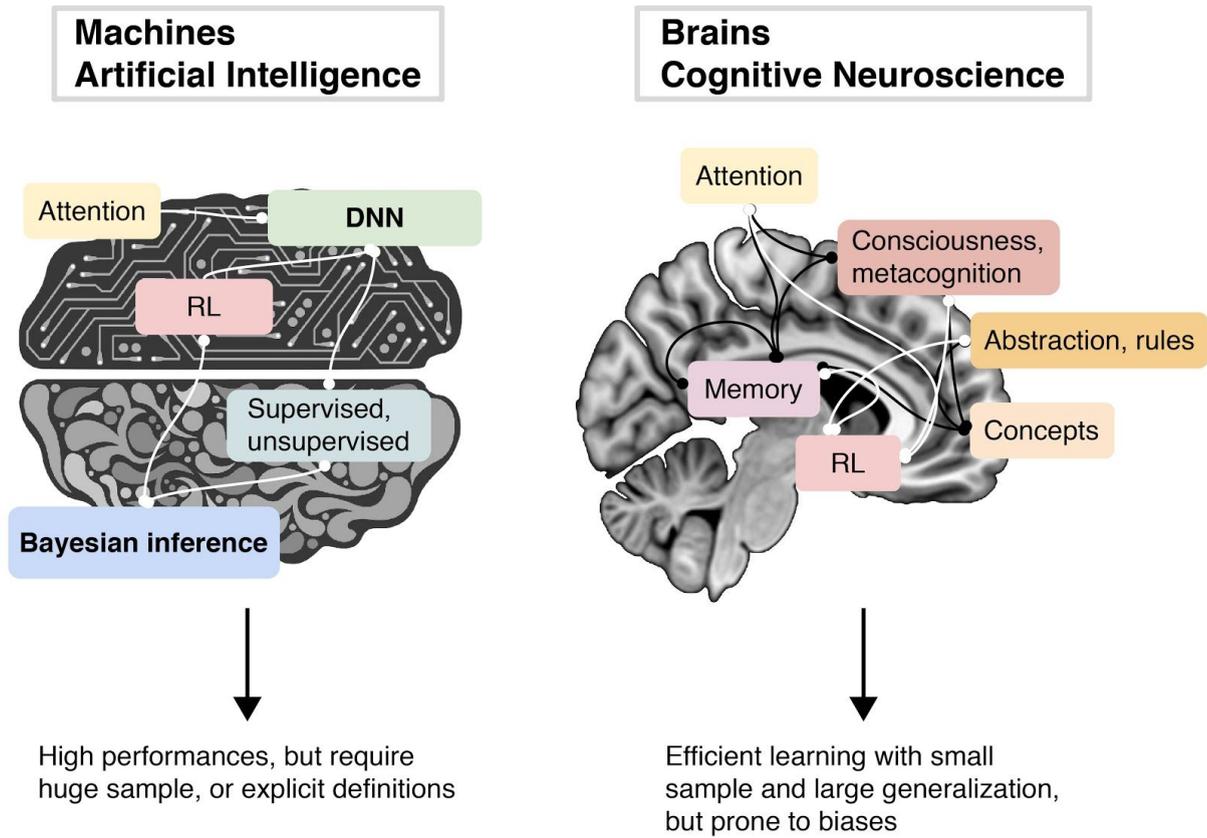

**Fig. 1: General schematic, solutions to complex problems in artificial intelligence and nature (brains).** Higher cognitive functions continuously interact between them and with reinforcement learning to drive generalization and learning from small sample.

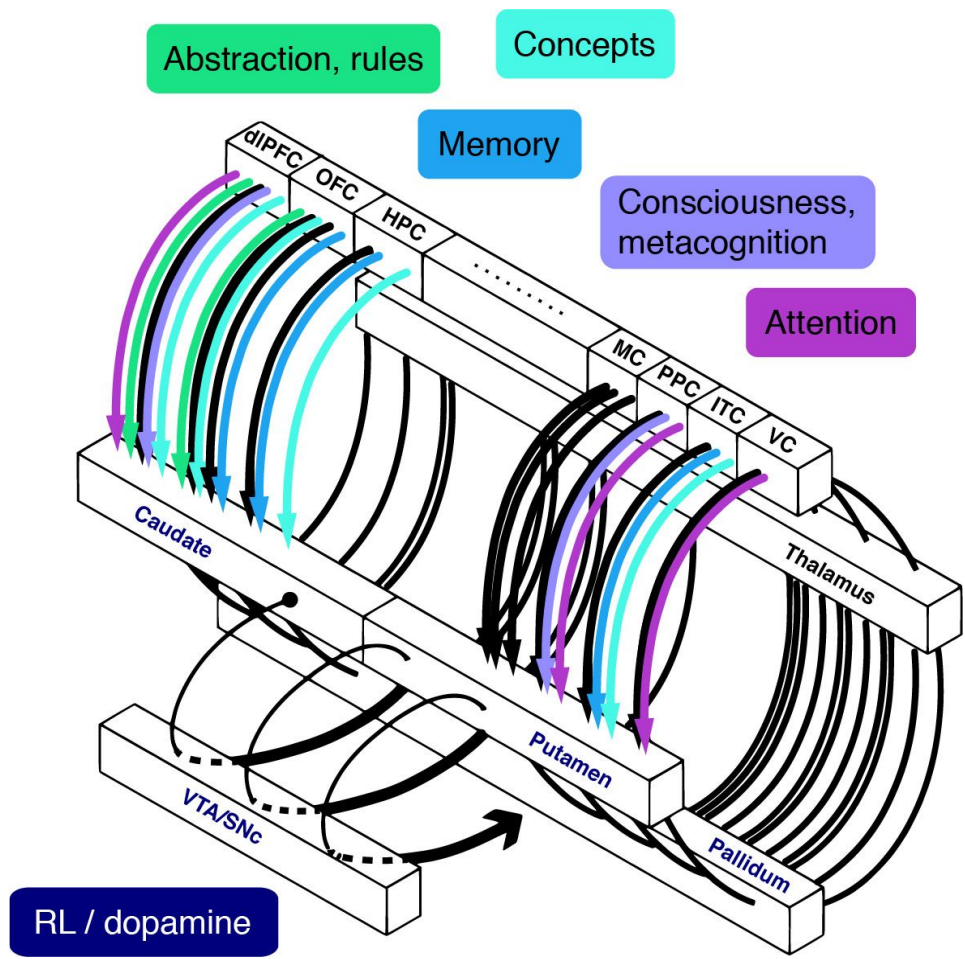

**Fig. 2: Winner-take-all parallel computing takes place in loops spanning the basal ganglia and neocortex.** Excitatory and inhibitory interactions between multiple loops formed by the basal ganglia, thalamus and cerebral cortical regions implement winner-take-all computations. The loop with the best representation for reinforcement learning and minimum reward prediction errors thus wins and all other loops are suppressed. These winner-take-all computations implement dimension reduction and feature selection, while being accelerated by high cognitive functions as follows. Attention executes feature selection rather than dimension reduction, with relatively low abstraction. Episodic memory, among different kinds of memories, represents feature selection in the time domain, and its abstraction level is relatively low. Conceptualization executes dimension reduction rather than feature selection, and its abstraction level is high. Metacognition does both dimension reduction and feature selection and its abstraction level is very high. Consciousness has the highest abstraction level and results in pure dimension reduction. dlPFC: dorsolateral prefrontal cortex, OFC: orbitofrontal cortex, HPC: hippocampal formation, MC: motor cortex, PPC: posterior parietal cortex, ITC: inferior temporal cortex, VC: visual cortex, VTA/SNc: ventral tegmental area / substantia nigra, RL: reinforcement learning. Figure modified from Haruno & Kawato [64].

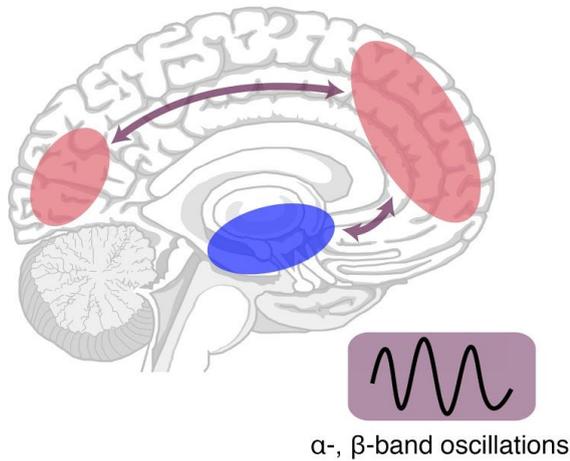
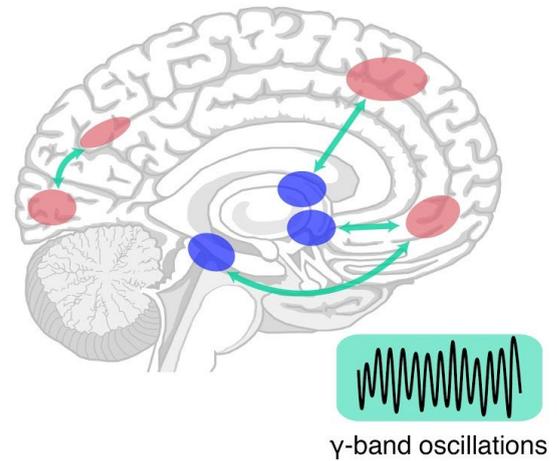
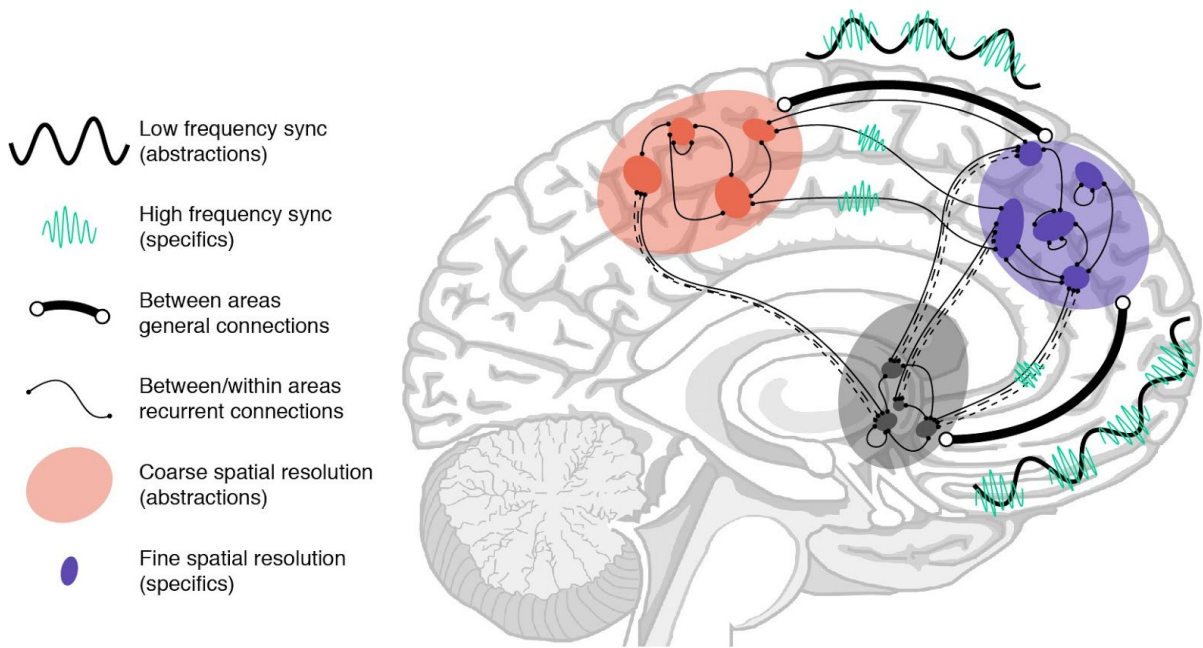

**Fig. 3: Different frequency modes in synchronization of neural activity represent coarse and fine dimension reduction and feature selection.** While low-frequency spatio-temporal modes contain many neurons and connections, high frequency modes contain small numbers of neurons and connections. Nonlinear dynamics interactions between low and high-frequency modes provide the computational means for fast parallel search of the optimal metarepresentation, corresponding smallest reward prediction errors (RPE), neurons and connections. When a low frequency mode is selected first, all high frequency modes contained within it are generally activated because low and

high frequency modes share common neurons and connections. Among the activated high-frequency modes, those with the highest correlations between meta-representations and RPE are further activated, and an optimal mode is thus selected. Consequently, dimension reduction with low-frequency synchronization and feature selection with high-frequency synchronization proceed together by closely interacting. Low frequency mode corresponds to dimension reduction such as principal component analysis (PCA) and high temperature in annealing. High frequency mode corresponds to feature selection such as L1-norm regularization or automatic relevance determination, and low temperature in annealing. Real or simulated annealing takes long time but brains cannot afford that. There exists no external control of temperature in the proposed interaction between different modes, and in a sense nonlinear brain dynamics analogously implement simulated annealing. With low signal to noise ratio, which is common in most learning problems, first an optimal low-frequency mode is activated because it contains many areas, neurons and connections. This increases the chances of correlation computations surviving high noise conditions. Then, high-frequency modes contained in it are generally activated, and correlations can be more reliably computed by constrained domains and general excitatory inputs to them. The selection of the optimal high-frequency mode can be executed more robustly. This interaction between low and high-frequency modes roughly implements annealing.